\documentclass[twoside,10pt,a4paper]{newFNLstyle}
\usepackage{graphics}
\usepackage{cite}

\begin{document}
\volnumpagesyear{0}{0}{000--000}{xxxx}
\dates{July 2003}{Oct 2003}{accepted date}
\pagenumbering{arabic}
\title{LANGEVIN APPROACH TO UNDERSTAND THE NOISE OF MICROWAVE TRANSISTORS}

\authorsone{FABIO PRINCIPATO, BERNARDO SPAGNOLO and GAETANO FERRANTE}
\affiliationone{INFM Unit$\grave{a}$ di Palermo and Dipartimento
di Fisica e Tecnologie Relative, Universit$\grave{a}$ di Palermo,
Viale delle Scienze, 90128 Palermo, ITALY}

\authorstwo{ALINA CADDEMI}
\affiliationtwo{INFM Unit$\grave{a}$ di Messina and Dipartimento
di Fisica della Materia e Tecnologie Fisiche Avanzate,
Universit$\grave{a}$ di Messina - Contrada Sperone, 31 - S. Agata,
98166 Messina, ITALY}


\maketitle

\markboth{Langevin Approach to Understand the Noise of Microwave
Transistors}{Principato et al.}

\pagestyle{myheadings}


\begin{abstract}
  A Langevin approach to understand the noise of microwave devices is presented.
  The device is represented by its equivalent circuit with the internal noise sources
  included as stochastic processes. From the circuit network analysis, a stochastic
  integral equation for the output voltage is derived and from its power spectrum the
  noise figure as a function of the operating frequency is obtained. The theoretical
  results have been compared with experimental data obtained by the characterization of an HEMT
  transistor series (NE20283A, by NEC) from 6 to 18 GHz at a low noise bias point.
  The reported procedure exhibits good accuracy, within the typical
  uncertainty range of any experimental determination. The approach allows to extract
  all the information required for understanding the noise performance of the device
  without any restriction on the statistics of the noise sources.
  The results show the relevant noise phenomena from a new angle.
\end{abstract}

\section{Introduction}
The noise performance of an active device up to millimeter-wave
frequencies plays a basic role in evaluating the suitability of
the latter for low-noise applications in advanced
telecommunication, nuclear instrumentation and radio astronomy
applications. This aspect is gaining increasing importance also
due to the implementation of nanometer scale devices where a
higher degree of interaction among the intrinsic noise sources is
expected to affect the device noise behavior \cite{Bal02}. So far,
the noise analysis approaches and the relevant modelling tools
employ the representation based on a second order statistics for
all the noise sources located inside a transistor \cite{Mars94}.
This allows to manage the analytical problems and leads to
satisfactory results within the 20 GHz range for device dimensions
not aggressively scaled, as for example FET's with gate length not
smaller than 0.2 $\mu$m.

In this paper we propose a new approach to the problem starting
from a general description of all noise sources and solving the
stochastic integral equation associated to the equivalent circuit
of the microwave device \cite{Spa01}. From the time waveform of
the output voltage we obtain the noise figure $F$ as a function of
the operating frequency. The noise figure, which is the link
between microscopic noise characteristics and macroscopic (i.e.
measurable) parameters \cite{Fri44}, is calculated by the ratio of
the spectral noise power density produced at the output of the
noisy device to the spectral noise power density produced at the
output of the ideal noiseless device. More specifically, our
calculation were performed for a noise source impedance of 50
$\Omega$, but the procedure can be applied for arbitrary impedance
value. We determine the device noise figure $F_{50}$ at different
discrete frequencies over a given operating range. Then, we
compare the values of $F_{50}$ calculated by means the proposed
method with data referring to characterization of an HEMT
transistor series (NE20283A, by NEC) from 6 to 18 GHz at a low
noise bias point \cite{Cad97}. The proposed procedure exhibits
fairly good accuracy, within the typical uncertainty range of any
experimental determination. It is worthwhile to note that for
Gaussian statistics the conventional CAD tools and our approach
coincide. However for non-Gaussian noise sources, i.e. when it is
important to know higher order statistics, our noise modelling is
appropriate. In fact with our approach we can use different
statistics of the noise sources and therefore obtain all the
moments of the output voltage as a function of the
cross-correlation functions of the noise sources. This technique
can thus be applied to extract all the information required for a
complete knowledge of the noise performance of the device without
any restriction on the statistics of the noise sources, including
also cross-correlated noise sources.

\section{HEMT Circuit Model}
High electron mobility transistors (HEMT) are GaAs or In-based
field-effect transistors for application at microwave and
millimeter wave frequencies. These devices began to set records
for noise performance since their introduction in the early '90s
and therefore represent key elements in all low-noise circuits.
The linear equivalent circuit which typically models the
small-signal performance of  HEMT's at microwave frequencies is
shown in Fig. 1. This circuit represents the chip device because
access inductances and parasitic intererelectrode capacitances are
not included.
\begin{figure}[htbp]
\centering{\resizebox{12cm}{!}{\includegraphics{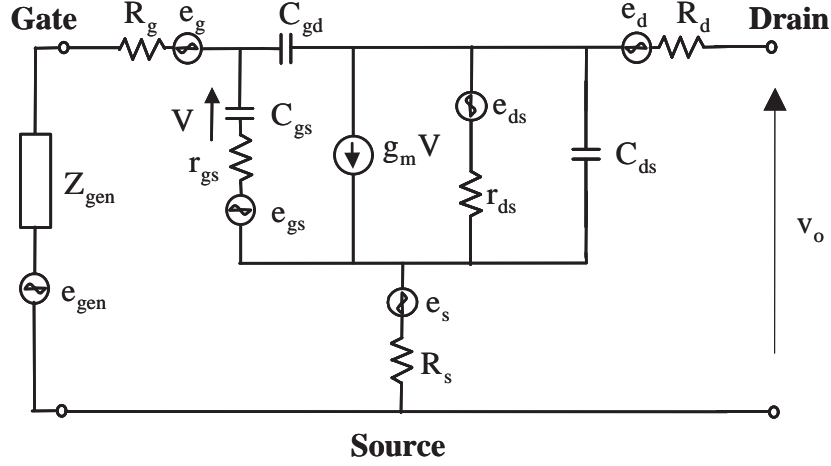}}}
\caption{Small-signal circuit model of a chip HEMT with noise
sources.}
\end{figure}
The values of the circuit elements are usually determined by
computer-aided modelling procedures which may follow either
optimization-based or direct extraction criteria \cite{Nie98,
Cad01,Cad97}. More recently, approaches based on artificial neural
networks have been successfully employed  \cite{Dev01}. We had
previously characterized an HEMT transistor series (NE20283A, by
NEC) over the $6-18$ GHz frequency range at the low noise bias
conditions suggested by the manufacturer ($V_{DS}=2.0\,V$,
$I_{DS}=10\, mA$). By microwave measurements, we determined both
the scattering and the noise parameters of the packaged device,
i.e. the device sealed within a ceramic enclosure equipped with
metallic leads. From the experimental data, we extracted a linear
noisy model for the whole structure by a variable decomposition
approach with a separate tuning for the channel noise source value
modelled by an equivalent noise temperature assigned to the
$r_{ds}$ resistor \cite{Pos89}. The values of the circuit elements
and the noise temperature $T_d$ so obtained are listed in Tab. 1.

\begin{table}[htbp]
\centering
\begin{tabular}{|l |c |c |c |c |c |c |c |c |c |c |} \hline
$r_{gs}$ & $r_{ds}$ & $R_g$ & $R_d$& $R_s$ &$C_{gs}$ &$C_{gd}$&
 $C_{ds}$ & $\tau$ &$g_{m0}$ & $T_d$ \\
$[\Omega$] & [$\Omega]$ & [$\Omega$] & [$\Omega$] & [$\Omega$] &
[$fF$] & [$fF$] & [$fF$] & [$ps$] &[$ mS$] & [$K$] \\ \hline
 1.2 & 274 & 2.0 & 0.67& 0.55 & 261& 20.0 & 34.6& 0.68&46.2 &2250
 \\\hline
\end{tabular}

\caption{Values of circuit elements of the chip HEMT model at
$V_{DS}=2.0\,V$ and $I_{DS}=10\,mA$.}
\end{table}

All the other resistors in the model contribute with thermal noise
typical of equilibrium conditions with device environment (i.e.
$300 K$). The comparison between the noise parameters measured and
modelled for the packaged device data was presented in
\cite{Cad97}. We here report for the reader's convenience the plot
of the noise figure in input matched condition in Fig. 2. To carry
out our analysis we employ the model structure representing the
chip device so eliminating the effect of the package found to be
lossless (i.e. not containing noise sources). Among all noise
parameters we use the $F_{50}$ noise figure as information test to
check the effectiveness of the procedure here presented.
Therefore, the new results are compared with $F_{50}$ data
referring to the chip as simulated via CAD by use of the above
circuit model.

\begin{figure}[htbp]
\centering{\resizebox{12cm}{!}{\includegraphics{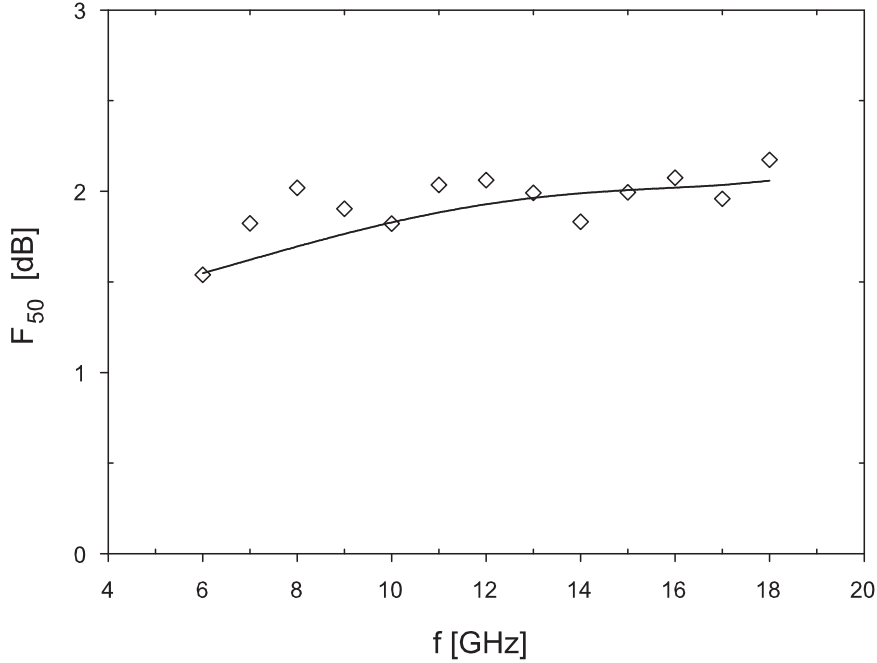}}}
\caption{Comparison between the measured noise figure $F_{50}$ of
the packaged HEMT transistor series NE20283A by NEC ($\diamond$)
and the $F_{50}$ simulated with CAD ($-$).}
\end{figure}

\section{General Noise Analysis by Langevin Approach}
For the sake of simplicity and to validate our theoretical
approach we suppose that all noise sources in the HEMT model are
Gaussian white noise of thermal origin with zero mean,
$\delta$-correlated and variance per unit of frequency bandwidth
equal to $\sigma^2=4kT_xR_x$, where $k$ is the Boltzmann constant,
$R_x$ the resistance and $T_x$ the equivalent temperature of the
resistor. No correlation is assumed among the noise sources. The
noise voltage generator $e_{gen}$ is the thermal noise associated
with the real part of the impedance $Z_{gen}=R_{gen}+j\,\omega
X_{gen}$ of the input termination at the reference temperature
$T_0=290 K$, which is the standard temperature to determine the
noise figure. The resistances $r_{gs}$ and $r_{ds}$ contribute to
noise by assigning equivalent temperature $T_g$ and $T_d$,
respectively. The temperature of the intrinsic gate-source
resistance $r_{gs}$ is assumed to be equal to the environment
temperature ($T_g=T_0$) \cite{Pos89}. In order to determine the
expression of the output noise voltage $v_o$  as a function of the
noise sources, the circuit network is analyzed in the Laplace
domain. So that we have the Laplace transform of the zero-state
(at $t=0$) response of the output voltage

\begin{equation}
 V_o(s)=\sum_{i=1}^{6}H_i(s)E_i(s),
 \label{laplace}
\end{equation}
where $s$ is a complex variable, $H_i(s)$ are the transfer
functions from the i-th noise generator to the output, and
$E_i(s)$ are the Laplace transforms of the voltage generators with
the following notation:  $1\rightarrow e_{gen}$, $2\rightarrow
e_{s}$, $3\rightarrow e_{d}$, $4\rightarrow e_{g}$, $5\rightarrow
e_{gs}$  and  $6\rightarrow e_{ds}$. In the equivalent circuit of
the HEMT, the small-signal $g_m$ transconductance depends on the
variable $s$ as $g_m=g_{m0} e^{-s \tau}$ , where $g_{m0}$ is the
dc value of the transconductance and $\tau$ is the transit time of
the carriers to diffuse from source to drain. Since the value of
$\tau$ is a fraction of picosecond, and we are interested in
investigating the frequency range below $20$ GHz, where
$|s=i\omega| \ll \frac{1}{\tau}$, we approximate the exponential
function as $e^{-s \tau}\simeq 1-s\tau +\frac{1}{2}(s\tau)^2$.
Consequently the $H_i(s)$ are rational functions of the complex
variable $s$. By anti-transforming (\ref{laplace}), we obtain the
zero-state response of the output voltage in the time domain which
assume the expression valid for $t>0$

\begin{equation}
v_o(t)=\sum_{i=1}^{6}\int_0^t h_i(t-t')\,e_i(t^{'})dt',
  \label{conv}
\end{equation}
where the function  $h_i(t)$ are the impulse responses
corresponding to the $H_i(s)$, which assume the expression

\begin{equation}
h_i(t)=h_{i0}\delta
(t)+h_{i1}e^{-\frac{t}{\tau_1}}+h_{i2}e^{-\frac{t}{\tau_2}}+h_{i3}e^{-\frac{t}{\tau_3}},\,
\, \,\,\,\,\,\,t>0. \label{rispimp}
\end{equation}
Here $\delta(t)$ is the Dirac $\delta$-function. The time
constants $\tau_i$ of the impulse response and the coefficients
$h_{ij}$ depend on the circuit parameters and are obtained from
the poles of the network functions $H_i(s)$(see Appendix A). For
the investigated model the following inequalities are satisfied

\begin{equation}
\tau_3 \ll \tau_1 \tau_2\, \mbox{ and     }\, \omega \ll
\frac{1}{\tau_3},
\end{equation}
thus the relation (\ref{rispimp}) can be approximated as:
\begin{equation}
h_i(t)\approx h_{i0}\delta
(t)+h_{i1}e^{-\frac{t}{\tau_1}}+h_{i2}e^{-\frac{t}{\tau_2}}.
\label{approxrispimp}
\end{equation}

By using Eqs. (\ref{conv}) and (\ref{approxrispimp}) the output
voltage of the HEMT model in the time domain can be expressed as
the following stochastic integral equation in the Ito sense

\begin{equation}
v_o(t)=\sum_{i=1}^{6}[h_{i0}\,e_i(t)+\int_0^t
h_i^{'}(t-t')\,e_i(t^{'})\,dt'],
\label{stochint}
\end{equation}
where the function $h_i^{'}$ contains only the last two terms of
the expression (\ref{approxrispimp}).The Eq.(6) is the main result
of our paper. It gives the stochastic output voltage of the HEMT
device as a function of the different stochastic processes which
represent the noise sources. By inserting the different statistics
of the noise sources in Eq.(6) we can extract all the moments of
the output voltage $v_o(t)$ required for a complete knowledge of
the noise performance when higher order statistics occurs in the
output signal of the devices. As an example we report in Appendix
B the expression of the second moment and the general expression
of the n\emph{th} moment of the output voltage as a function of
the cross-correlation functions of the noise sources. As a first
step of investigations and to validate our stochastic approach we
consider all noise generators $e_i(t)$ present in the model as
uncorrelated Gaussian white noises, with the usual statistical
properties: $\langle e_i(t)\rangle=0$ and $\langle
e_i(t)\,e_i(t^{'})\rangle=\sigma_i^2 \delta(t-t^{'})$. Here
$\sigma_i^2$ is the variance per unit of frequency bandwidth of
the i-th generator. Therefore the time waveform of the output
voltage can be obtained by numerical integration of Eq.
(\ref{stochint}). In Eq. (\ref{stochint}) the terms
$e_i(t)\,dt=dW_i(t)$ are Wiener processes associated with the
device noise sources \cite{Gar85}. The numerical simulation of
Eq.(\ref{stochint}) is performed by choosing the time step $\Delta
t=min\{\tau_1,\tau_2\}/10$, in order to prevent aliasing. At each
temporal step $t_n=n\,\Delta t$, by using the algorithm for
Gaussian random variables \cite{Pre92}, the Wiener increments are
calculated as $\Delta W_i(t_n)=\sigma_i \xi(t_n)\sqrt{\Delta t}$.
The digital implementation of the impulse response
(\ref{approxrispimp}) allows to calculate the output voltage by
the discrete-time convolution between the sequences $h^{'}_i(t_n)$
and the noise sources $e_i(t_n)$

\begin{equation}
\sum_{i=1}^{6}\sum_{k=0}^{n}h^{'}_i(t_{n-k}) \Delta W_i(t_k).
\end{equation}
The duration of the sequence $h^{'}_i(t_n)$ increases whit $n$,
therefore we consider only the first $n_{max}$ terms of the
sequence $h^{'}_i(t_n)$ to save computation time. That is

\begin{equation}
v_o(t_n)=\sum_{i=1}^{6}\sum_{k=n-n_{max}}^{n}[h_{i0}\,e_i(t_n)+h^{'}_i(t_{n-k})\,
\Delta W_i(t_k)].
 \label{discrconv}
\end{equation}
The time discretization and the finite duration of the impulse
response (FIR) introduce an error. In order to evaluate this error
the z-transform of the sequence
$h_i(t_0),h_i(t_1),h_i(t_2),...,h_i(t_{n_{max}})$  is considered

\begin{equation}
H_{di}(z)=\sum_{n=0}^{n_{max}}h_{i}(t_n)\,z^{-n}.
\end{equation}
The error is evaluated by comparing the module of the Fourier
transform of the continuous transfer function, i.e.

\begin{equation}
H_{i}(i\omega)=h_{i0}+\frac{h_{i1}}{1/\tau_1+i\omega}+\frac{h_{i2}}{1/\tau_2+i\omega},
\end{equation}
with the Fourier transform of the sequence $h_i(t_n)$, i.e.
$H_{di}(z=e^{i\omega \Delta t}$), at different values of $n_{max}$
and for each generator $e_i$. The comparison between the
continuous and the discrete transfer function is shown in Fig. 3,
where we report the normalized Fourier transform of the transfer
function versus the normalized frequency. In this figure we use
$n_{max}=m\times max\{\tau_1,\tau_2\}$, with $m$ integer, and
$f_s=1/(2\,\Delta t)$ is the Nyquist frequency. As $m$ increases,
the difference between continuous and the approximate impulse
response becomes negligible. At $m=6$ the error introduced by the
FIR impulse response are less than 1\%. We use this value of $m$
in our simulations.
\begin{figure}[htbp]
\centering{\resizebox{10cm}{!}{\includegraphics{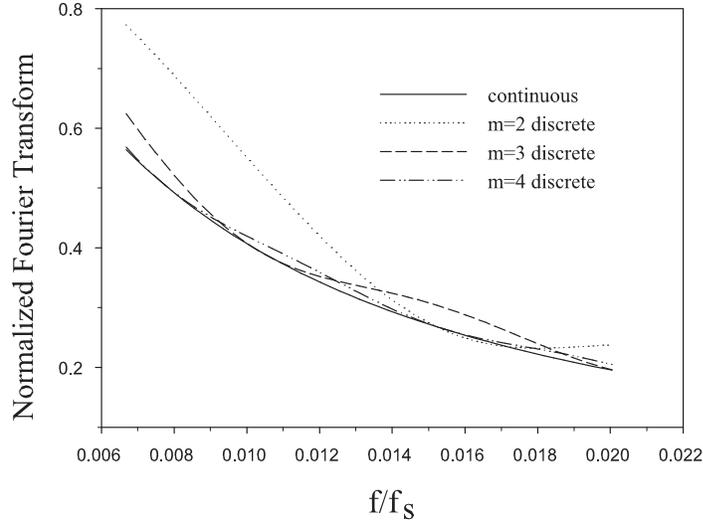}}}
\caption{Normalized module of the Fourier transform of the
transfer function from the input generator to voltage output for
the continuous and discrete impulse response and for different
values of the integer $m$. In the x-axis the frequency is
normalized with respect to the Nyquist frequency and the values
are restricted to the $6-18$ GHz frequency band.}
\end{figure}

\section{Results}
The samples of the output noise voltage waveforms were calculated
for an input termination $50\,\Omega$ from Eq. (\ref{discrconv})
by using C-language routines. To validate the approach proposed in
this paper we compare the theoretical results of the noise figure
$F_{50}$ with that obtained by CAD analysis in the 6-18 GHz
frequency range. The temporal noise sequence was generated by
using a time step $\Delta t$ of $5.57\times10^{-13}$ sec and
Gaussian random number generator. The Nyquist frequency is 0.898
THz, thus $N=10^{5}$ points are needed at least to investigate the
microwave frequency range. The noise figure is calculated by using
the following relation \cite{Pri99}:
\begin{equation}
F_{50}=\frac{p_{on}(f)}{p_{off}(f)}
\label{F50}
\end{equation}
where $p_{on}(f)$ is the power spectral density of the output
voltage at the frequency $f$ when all the noise sources are
activated and $p_{off}(f)$ is the power spectral density obtained
from the noise source of the input termination only, both in
$V^2/Hz$ units. In order to improve the estimation of the spectral
power density given by (\ref{F50}) at a given value of frequency
$\bar{f}$, we use the mean value of the noise power evaluated
within the frequency bandwidth $BW$ around $\bar{f}$

\begin{equation}
\langle p_x(\bar{f})\rangle=\frac{1}{BW}
\int^{\bar{f}+BW/2}_{\bar{f}-BW/2}p_x(f)\,df.
\end{equation}
The output noise waveforms of the HEMT model, with ($v_{on}$) and
without ($v_{off}$) the internal noise sources, over a sample time
interval of 20 ns are reported in Fig. 4. The total time period
used for calculation is  $N\,\Delta t$, where $N$  is usually
chosen to be $1\times 10^5$.

\begin{figure}[htbp]
\centering{\resizebox{13cm}{!}{\includegraphics{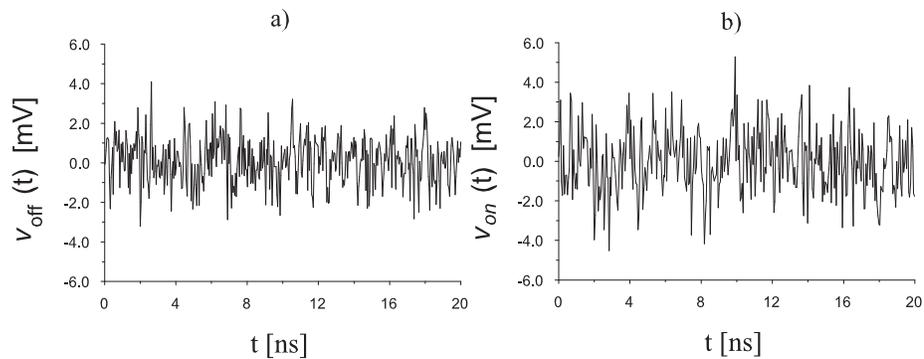}}}
\caption{Output noise waveforms of the HEMT model without
$v_{off}$ (a) and with $v_{on}$ (b) internal noise sources.}
\end{figure}
By using  $BW=90$ MHz for noise power estimation, we calculate the
$F_{50}$ as a function of the operating frequency in the 6-18 GHz
range. In Fig. 5 we compare the noise figure behavior obtained
with our approach with that obtained by computer-aided analysis of
the model.

\begin{figure}[htbp]
\centering{\resizebox{12cm}{!}{\includegraphics{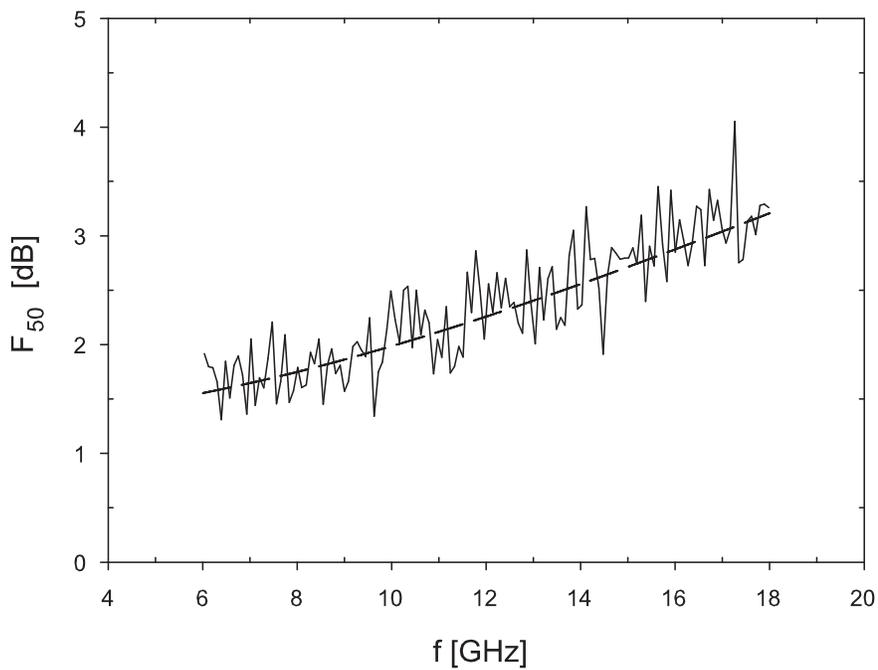}}}
\caption{Noise figure $F_{50}$ calculated with Langevin approach
and by using $BW=90$ MHz for noise power estimation (---) compared
with $F_{50}$ obtained by computer-aided analysis of the model
(--\,--\,).}
\end{figure}
Good agreement is exhibited by the two noise figure curves. We use
100 realizations of time waveforms to determine each noise figure
behavior over the given frequency range. The results obtained with
the stochastic approach may be improved by averaging over a
greater number of realizations but with more computer time
consuming. We note that the noise figure behavior obtained with
the stochastic approach could be thought as the output of a noise
figure measuring instrument where averaging and smoothing options
are available to minimize display jitter. The procedure here
described can be used with any input termination value to obtain
more than one noise figure value. By choosing four different
termination values, the extraction of all the standard noise
parameters can be performed by using the equation

\begin{equation}
F(\Gamma_{gen})=F_0+4\,r_n\,\frac{|\Gamma_{gen}-\Gamma_o|^2}{|1+\Gamma_o|^2\,(1-|\Gamma_{gen}|^2)}
\label{}
\end{equation}
where  $\Gamma_{gen}$ is the generic termination value expressed
in a reflection coefficient form and $F_0$, $\Gamma_o$ (magnitude
and angle) and $r_n$ are the four noise parameters usually
determined by complex experimental procedures \cite{Mar94}.

By calculating the noise parameters vs. frequency, a complete
knowledge of the noise behavior of any two-port is obtained from a
circuit point of view. The noise parameters can be employed either
for extraction of device intrinsic noise sources or for CAD of
low-noise circuits.

\section{Conclusions} We have presented a noise analysis
procedure for microwave transistors. This procedure is based on
Langevin approach and takes the general statistical properties of
the internal noise sources into account. The device is represented
by its equivalent circuit with noise generators to a chip level.
The noise figure at a given input termination is determined by
first integrating the noise voltage equations in the time domain
and by transforming them into spectral power densities over the
frequency range of interest. The results are compared with the
simulated data, obtained with a commercial CAD, of the chip model
of an HEMT device series (NE20283A, by NEC) from 6 to 18 GHz at a
low noise bias point. The proposed procedure exhibits good
accuracy within the typical uncertainty range of any experimental
determination. Our Langevin approach gives the same results of
noise analysis approaches based on second order statistics. But
when higher order statistics becomes relevant to the noise
analysis our approach is suitable. Moreover the cross-correlation
functions between the noise sources of the device are explicitly
included in our theoretical approach (see Appendix B), while in
conventional CAD methods their are inserted in a heuristic way.
Such a general approach can be extended to analyze noise in other
devices without restrictions as to the frequency range, by
specifying the device equivalent circuit and the statistical
properties of the internal noise sources.

\section*{Acknowledgements}
The authors are grateful to Dr. D. Valenti for helpful discussions
concerning simulation. This work was supported by the National
Institute for the Physics of Matter (I.N.F.M.), by INTAS Grant
01-450 and by Italian Ministry of University and Research (MIUR).

\appendix

\section{} Because of the topological complexity of the HEMT model
network it is not possible to obtain the expression of the time
constants $\tau_i$ in a closed form. Indeed the $\tau_i$ are
obtained from the poles of the transfer function $H_i(s)$, which
are the roots of the third-order algebraic equation

\begin{eqnarray}
\nonumber ~& ~&2\,(1+(C_{{{\it gs}}}r_{{{\it gs}}}+C_{{{\it
gd}}}r_{{{\it ds}}}+C_{{{\it ds}}}r_{{{\it ds}}}+C_{{{\it
gs}}}R_{{g}}+C_{{{\it gd}}}R_{{g}}+R_{{{ \it gen}}}C_{{{\it
gd}}}+R_{{{\it gen}}}C_{{{\it gs}}} \\
\nonumber ~& ~&+g_{{{\it m0}}}r _{{{\it ds}}}C_{{{\it
gd}}}R_{{g}}+R_{{s}}C_{{{\it gs}}}+g_{{{\it m0}} }R_{{s}}r_{{{\it
ds}}}C_{{{\it gd}}}+R_{{s}}C_{{{\it gd}}}+g_{{{\it m0 }}}r_{{{\it
ds}}}C_{{{\it gd}}}R_{{{\it gen}}})\,s\\
\nonumber ~& ~&+(C_{{{\it ds}}}r_{{{\it ds}}}C_{{{\it
gs}}}r_{{{\it gs}}}+C_{{{\it ds}} }r_{{{\it ds}}}C_{{{\it
gd}}}R_{{g}}+C_{{{\it gd}}}r_{{{\it ds}}}C_{{{ \it
gs}}}R_{{g}}+C_{{{\it ds}}}r_{{{\it ds}}}C_{{{\it gs}}}R_{{g}}+C_{
{{\it gs}}}r_{{{\it gs}}}C_{{{\it gd}}}R_{{g}}
\\
\nonumber ~& ~&-g_{{{\it m0}}}r_{{{\it ds}}}\tau\,C_{{{\it
gd}}}R_{{g}}+r_{{{\it ds}}}C_{{{\it gs}}}r_{{{\it gs}}}C_{{{\it
gd}}}+R_{{s}}C_{{{\it ds}}}r_{{{\it ds}}}C_{{{\it gd}}}+
R_{{s}}r_{{{\it ds}}}C_{{{\it gs}}}C_{{{\it gd}}}+R_{{s}}C_{{{\it
ds}} }r_{{{\it ds}}}C_{{{\it gs}}}\\
\nonumber ~& ~&+r_{{{\it gs}}}R_{{s}}C_{{{\it gs}}}C_{{{ \it
gd}}}-g_{{{\it m0}}}R_{{s}}r_{{{\it ds}}}\tau\,C_{{{\it gd}}}+C_{{
{\it gd}}}r_{{{\it ds}}}R_{{{\it gen}}}C_{{{\it gs}}}+C_{{{\it
ds}}}r_ {{{\it ds}}}R_{{{\it gen}}}C_{{{\it gs}}} \\
\nonumber ~& ~&+C_{{{\it ds}}}r_{{{\it ds}}} R_{{{\it
gen}}}C_{{{\it gd}}}+C_{{{\it gs}}}r_{{{\it gs}}}C_{{{\it gd}
}}R_{{{\it gen}}}-g_{{{\it m0}}}r_{{{\it ds}}}\tau\,C_{{{\it
gd}}}R_{{ {\it gen}}})\,s^2\,) \\
\nonumber ~& ~&+(C_{{{\it gd}}}r_{{{\it ds}}}\left (g_{{{\it
m0}}}{\tau}^{2}+2\,r_{{{ \it gs}}}C_{{{\it gs}}}C_{{{\it
ds}}}\right )\left (R_{{g}}+R_{{s}}+R_ {{{\it gen}}}\right ))\,s^3=0 .\\
\end{eqnarray}

\section{}

In the HEMT model the only plausible cross-correlated noise
sources are the gate-source $e_{gs}$ and the drain-source $e_{ds}$
\cite{Pos89}. Do to this hypothesis we obtain from Eq. (6) the
general expression for the second moment of the output voltage
$v_0(t)$ as a function of the cross-correlation function between
the $e_{gs}$ and the $e_{ds}$ noise sources

\begin{eqnarray}
 \nonumber ~& ~&\langle v_0^2(t) \rangle=\sum_{k=0}^6 h_{i0}^2 \sigma_i+2\,h_{50}\,h_{60}\langle e_5(t)\,e_6(t')
 \rangle+\sum_{k=0}^6 \int_0^t h{'}_i^{2}(t-t{'}) \sigma_i\,dt{'}\\
 \nonumber ~& ~&+2\,\int_0^t \int_0^t h{'}_5(t-t^{'})\,h{'}_6(t-t{''})\langle e_5(t')\,e_6(t'') \rangle
 dt'\,dt''+ \,\sum_{k=0}^6 h_{i0} h_{i}^{'}(0) \sigma_i\\
 \nonumber ~& ~&+h_{50}\int_0^t h'_5(t-t')\langle e_5(t)\,e_6(t')
 \rangle dt'+h_{60}\int_0^t h'_6(t-t')\langle e_6(t)\,e_5(t') \rangle dt,\\
 \end{eqnarray}

\noindent
 where we use the property of the Dirac
$\delta$-function. The general expression of the moments of
$v_0(t)$ with the same above-mentioned  hypothesis is

\begin{eqnarray}
 \nonumber ~& ~&\langle v_0^n(t) \rangle=\sum_{k=0}^n
 \sum_{g=0}^{(n-k)} \sum_{j=0}^{(n-k-g)} \sum_{l=0}^{(n-k-g-j)}
\sum_{p=0}^{(n-k-g-j-l)}
 { n \choose k } { n-k \choose g } { n-k-g \choose j } \\
 \nonumber ~& ~& {n-k-g-j \choose l }
 {n-k-g-j-l \choose p } \langle \sum_{m=0}^k { k \choose m }
 \left [ \int_0^t h'_6(t-t') e_6(t')dt' \right ]^m \\
 \nonumber ~& ~& \cdot(h_{60} e_6(t))^{k-m}
 \sum_{q=0}^g { g \choose q } \left [ \int_0^t h'_5(t-t') e_5(t')dt' \right ]^q
 (h_{50} e_5(t))^{g-q} \rangle  \\
 \nonumber ~& ~& \cdot \sum_{r=0}^j { j \choose r} \langle
 \left [ \int_0^t h'_4(t-t') e_4(t')dt' \right ]^{r}(h_{40} e_4(t))^{j-r} \rangle
 \\
 \nonumber ~& ~&  \cdot \sum_{s=0}^l { l \choose s } \langle
 \left [ \int_0^t h'_3(t-t') e_3(t')dt' \right ]^{s}(h_{30} e_3(t))^{l-s}
 \rangle \\
 \nonumber ~& ~&
  \cdot \sum_{u=0}^p { p \choose u }
 \langle \left [ \int_0^t h'_2(t-t') e_2(t')dt' \right ]^{u}(h_{20} e_2(t))^{p-u}
 \rangle \\
 \nonumber ~& ~& \cdot \sum_{z=0}^{(n-k-g-j-l-p)} { n-k-g-j-l-p \choose z
 }\\
\nonumber ~& ~& \cdot \langle \left [ \int_0^t h'_1(t-t')
e_1(t')dt' \right ]^{z}
(h_{10} e_1(t))^{(n-k-g-j-l-p)-z}\rangle \\
 \end{eqnarray}

\end{document}